\documentclass[twocolumn]{aastex61}

\renewcommand{\_}[1]{_\mathrm{#1}}

\newcommand{\NBa}{\textit{NB711}}
\newcommand{\NBb}{\textit{NB671}}
\newcommand{\B}{\textit{B}}
\newcommand{\R}{\textit{R}}
\newcommand{\ip}{\textit{i'}}
\newcommand{\z}{\textit{z'}}

\received{}
\revised{}
\accepted{}
\submitjournal{ApJ}

\begin{document}

\title{Active Galactic Nucleus Environments and Feedback to Neighboring Galaxies at $z\sim5$ Probed by Lyman-Alpha Emitters\footnote{Based on data collected at Subaru Telescope, which is operated by the National Astronomical Observatory of Japan.}}
\shorttitle{AGN environments and feedback}

\author{Satoshi Kikuta}
\affil{Department of Astronomical Science, SOKENDAI (The Graduate University for Advanced Studies), \\
Osawa, Mitaka, Tokyo 181-8588, Japan}
\affil{National Astronomical Observatory of Japan, Osawa, Mitaka, Tokyo 181-8588, Japan}
\email{satoshi.kikuta@nao.ac.jp}
\shortauthors{Kikuta et al.}
\author{Masatoshi Imanishi}
\affil{Department of Astronomical Science, SOKENDAI (The Graduate University for Advanced Studies), \\
Osawa, Mitaka, Tokyo 181-8588, Japan}
\affil{National Astronomical Observatory of Japan, Osawa, Mitaka, Tokyo 181-8588, Japan}
\author{Yoshiki Matsuoka}
\affil{Department of Astronomical Science, SOKENDAI (The Graduate University for Advanced Studies), \\
Osawa, Mitaka, Tokyo 181-8588, Japan}
\affil{National Astronomical Observatory of Japan, Osawa, Mitaka, Tokyo 181-8588, Japan}
\affil{Research Center for Space and Cosmic Evolution, Ehime University, Matsuyama, Ehime 790-8577, Japan.}
\author{Yuichi Matsuda}
\affil{Department of Astronomical Science, SOKENDAI (The Graduate University for Advanced Studies), \\
Osawa, Mitaka, Tokyo 181-8588, Japan}
\affil{National Astronomical Observatory of Japan, Osawa, Mitaka, Tokyo 181-8588, Japan}
\author{Kazuhiro Shimasaku}
\affil{Department of Astronomy, Graduate School of Science, The University of Tokyo, Tokyo 113-0033, Japan}
\affil{Research Center for the Early Universe, The Graduate School of Science, The University of Tokyo, Tokyo 113-0033, Japan}
\author{Fumiaki Nakata}
\affil{Subaru Telescope, 650 North A’ohoku Place, Hilo, HI 96720, USA}

\correspondingauthor{Satoshi Kikuta}

\begin{abstract}
Active galactic nuclei (AGNs) in the high-redshift Universe are thought to reside 
in overdense environments. However, recent works provide controversial results 
partly due to the use of different techniques and possible suppression of nearby 
galaxy formation by AGN feedback. 
We conducted deep and wide-field imaging observations with the Suprime-Cam on the 
Subaru telescope and searched for Lyman-alpha emitters (LAEs) around two QSOs (quasi-stellar objects) at 
$z\sim4.9$ and a radio galaxy at $z\sim4.5$ by using narrow-band filters 
to address these issues more robustly.
In the QSO fields, we obtained additional broad-band images to select Lyman-break 
galaxies (LBGs) at $z\sim5$ for comparison. 
We constructed a photometric sample of 301 LAEs and 170 LBGs in total. 
A wide field of view (34\arcmin$\times$27\arcmin, corresponding to 80$\times$60 
comoving Mpc$^2$) of the Suprime-Cam enabled us to probe galaxies in the immediate 
vicinities of the AGNs and in the blank fields simultaneously and compare various 
properties of them in a consistent manner. 
The two QSOs are located near local density peaks ($<2\sigma$) 
and one of the QSOs has a close companion LAE with projected separation of 80 physical kpc. The radio galaxy is found to be near a void of LAEs. 
The number densities of LAEs/LGBs in a larger spatial scale around the AGNs are not 
significantly different from those in blank fields. No sign of feedback is found 
down to $L\_{Ly\alpha}\sim10^{41.8}\mathrm{~erg~s^{-1}}$. Our results suggest that 
high-redshift AGNs are not associated with extreme galaxy overdensity 
and that this cannot be attributed to the effect of AGN feedback. 
\end{abstract}


\section{Introduction} \label{sec:intro}
Supermassive black holes (SMBHs) at the center of galaxies provide us 
various insights into key physics of galaxy formation and evolution.
The correlation between the properties of SMBHs and those of the host galaxies 
such as masses and velocity dispersions of spheroidal components
\citep{Magorrian1998,Ferrarese2000} indicates that active galactic nuclei (AGNs; rapidly growing SMBHs) 
and star formation activities are
physically connected and that AGNs play a crucial role in galaxy formation. 
However, details of their formation and growth history still remain largely unknown. To date, SMBHs with BH masses 
in excess of $10^9~M_\odot$ are known to already exist at the 
very beginning of the Universe ($z>6$; e.g., \citealt{Mortlock2011,Wu2015}). 
Where and how these SMBHs are formed and evolved in pace with their host galaxies is a fundamental question to elucidate galaxy evolution.

It is expected that high-redshift SMBHs should reside more 
frequently in highly biased regions of the Universe where the dark matter 
and galaxies are overly clustered \citep[e.g.,][]{Volonteri2006}. 
Locating such overdense regions at high-redshift Universe, 
the so-called ``protoclusters" (\citealt{Shimasaku2003,Matsuda2010,Toshikawa2012}, 
see \citealt{Overzier2016} 
for a recent review) has a potential to provide us 
opportunities to probe how environmental differences in galaxy evolution 
observed at the local Universe \citep{Dressler1980a,Cappellari2011} 
are established.
Since searching for protoclusters without signposts like AGNs is 
difficult because of their rarity, probing AGN environments 
gives us clues to environmental effects on galaxy evolution 
as well as growth history of SMBHs. 

Environments of high-redshift quasi-stellar objects (QSOs; luminous type 1 AGNs) and 
radio galaxies (RGs, radio-loud type 2 AGNs) have been 
extensively studied with the motivation described above. 
Radio-loud AGNs (RGs and radio-loud QSOs) are typically found in 
overdense regions using various methods to locate galaxies around them, including 
the Lyman break technique, narrow-band imaging, and Spitzer/Infrared Array Camera (IRAC) color-selection
\citep{Zheng2006,Venemans2007,Matsuda2009,Wylezalek2013}. On the other hand, 
the situation is different for (radio-quiet) QSOs. 
While measurements of QSO clustering shows that they should reside in massive 
dark matter halos ($M\_h > 10^{12} ~\mathrm{M_\odot}$) up to $z<4$ 
(\citealt{Shen2007,Trainor2012,Garcia-Vergara2017}, 
see also \citealt{Eftekharzadeh2015}), 
many authors found both galaxy overdensity associated with QSOs 
\citep{Utsumi2010,Capak2011,Falder2011,Swinbank2012,Husband2013,Morselli2014} 
and normal or underdensity around QSOs 
\citep{Francis2004,Kashikawa2007,Kim2009,Banados2013,Simpson2014,Adams2015,Mazzucchelli2016}, 
again using various technique to identify galaxies in the QSO fields.
Even in fields around intermediate-redshift QSO multiples, their environments are not 
always rich \citep{Boris2007,Farina2013,Sandrinelli2014}, 
though \citet{Hennawi2015} found an overdensity around a quasar quartet.

Several authors suggest that strong UV radiation from AGNs can suppress 
the formation of low-mass galaxies around them by heating and photoevaporating 
their gas \citep{Kashikawa2007,Utsumi2010,Bruns2012}, and thereby dilute 
any sign of overdensity and mitigate the discrepancy observed to date. 
The deficit of \ion{H}{1} gas with column density 
N$_{\rm H}$ $<$ 10$^{17}$ cm$^{-2}$ (Lyman-alpha forest)  
within a few to several physical Mpc (pMpc) from QSOs has been well known as the QSO
proximity effect \citep{Bechtold1994,Calverley2011}. However, it is unclear 
whether this QSO's radiative feedback is indeed strong enough to suppress 
the formation of neighboring low-mass galaxies 
(N$_{\rm H}$ $>$ 10$^{20}$~cm$^{-2}$). 
On the other hand, \citet{Cantalupo2012} claimed that fluorescently illuminated gas 
around a hyperluminous QSO can boost Ly$\alpha$ luminosity of galaxies around the QSO. 

Mixed results about QSO environments obtained so far can be partly due to 
the use of different observational techniques, survey depths, and field coverages, 
as well as various feedback effects. 
Many studies used the Lyman break technique to identify high-redshift galaxies 
around QSOs. However, this technique sample galaxies from wide redshift range of, 
say, $\Delta z \sim 1$ \citep{Yoshida2006}. At $z\sim5$, this corresponds 
to $\sim100$ pMpc. This scale is far larger than the expected size of 
the largest protocluster and known QSO proximity, 
hence galaxies selected by this method contain many 
foreground and background, physically unrelated galaxies. 
Moreover, even if AGNs have associated structures of a scale of a few to 
several pMpc, or if AGNs affect surrounding galaxies within this 
scale, it should be smeared out by the projection effect. 
If one wants to study AGN environments and feedback to their neighbors, 
it is particularly important to pick up galaxies within the AGN's proximity 
in both tangential and radial (redshift) directions.

Wide-field, narrow-band imaging observations are currently the best way to securely investigate 
AGN environments and feedback unless a large spectroscopic 
sample is available. Narrow-band filters have full widths at half maximum (FWHMs) of 
$\lesssim100${\AA} and they enable us to select line emitting galaxies 
from a narrow redshift range of $\Delta z\lesssim0.1$. 
In particular, galaxies whose redshifted Ly$\alpha$ emission fall inside a narrow-band 
filter, the so-called Lyman-alpha emitters (LAEs), are commonly found at $z>2$. 
At $z\sim5$, $\Delta z \sim0.1$ corresponds to $\sim10$ pMpc. 
This scale is sufficiently small to detect protoclusters of this redshift 
\citep{Chiang2013} 
and matches measured QSO's proximity sizes \citep{Calverley2011}. 
With narrow-band filters, we can select LAEs within the AGN’s proximity in 
radial (redshift) direction, by minimizing the contamination from physically 
unrelated foreground and background galaxies. 
Furthermore, majority of LAEs are known to have low stellar mass
\citep{Gawiser2006,Finkelstein2007,Ono2010a,Ono2010} and therefore more susceptible 
to AGN feedback than massive galaxies. Thus, LAEs are best suited to observationally 
scrutinize whether and how the properties of low-mass galaxies around AGNs are altered 
compared to general fields, by overcoming 
the uncertainties in previous works.

In this paper, we present the results of our deep and wide-field observations 
around two QSOs at $z\sim4.9$ and one RG at $z\sim4.5$ using the Suprime-Cam 
\citep[S-Cam,][]{Miyazaki2002} on the Subaru telescope. Narrow-band filters well covering 
the proximity of these AGNs in the radial direction and the wide field of view (FoV) of 
S-Cam ($34\arcmin\times27\arcmin$, or $13\times10$ pMpc at $z=5$) enables us 
to detect galaxies within and outside of the AGN's proximity 
simultaneously, as opposed to previous narrow-band 
studies with smaller FoVs 
\citep{Swinbank2012,Banados2013,Mazzucchelli2016}. 
We try to identify the effect of AGN feedback by comparing 
the luminosity functions (LFs) within and outside of the AGN proximity, 
because radiative feedback affects galaxies differently depending on their mass. 

The structure of this paper is as follows: in Section \ref{sec:obs}, we describe 
our imaging observations and data reduction. 
In Section \ref{sec:sample} we present our photometric selection of LAEs and LBGs 
and subsequent analyses. 
We show our main results and discuss them in Section \ref{sec:results} and 
present the summary and conclusion in Section \ref{sec:summary}.
Throughout this paper, we use the AB magnitude system and adopt a $\Lambda$CDM cosmology, with $H_0 =71 \mathrm{~km~s^{-1}~Mpc^{-1}}$, $\Omega\_M=0.27$, $\Omega_\Lambda=0.73$ \citep{Komatsu2009}.

\section{observations and data reduction} \label{sec:obs}

\begin{figure}
   \plotone{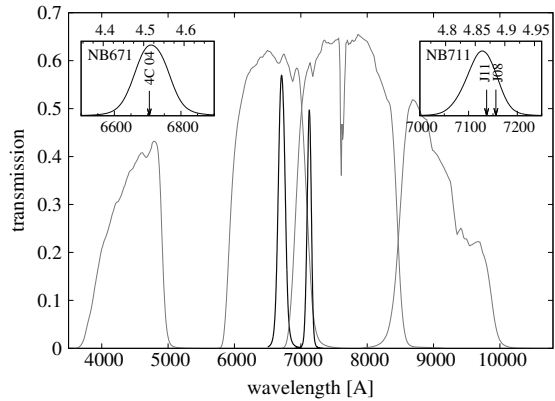}
 \caption{Transmission curves of filters we used. Thin curves indicate transmission of broad-band filters (from left to right, \B, \R, \ip, and \z). Thick curves on the left (right) shows that of \NBa (\NBb). Insets show enlarged transmission curves of narrow-band filters. The upper axis shows corresponding Ly$\alpha$ redshift. The redshift of AGNs are marked with arrows. 
\label{fig:trans}}
\end{figure}

\floattable
\begin{deluxetable}{ccCCccccc}[htb!]
\tablecaption{Information about our targets and observations. \label{tab:obs}}
\tablehead{
\colhead{Object name} &
\colhead{Redshift\tablenotemark{a}} &
\colhead{$\log M\_{BH}$\tablenotemark{a}} & 
\colhead{$\log L\_{bol}$\tablenotemark{a}} &
\multicolumn{5}{c}{Exposure time (sec) and seeing\tablenotemark{b}} \\
\colhead{} & \colhead{} &
\colhead{($\mathrm{M_\odot}$)} & \colhead{($\mathrm{erg~s^{-1}}$)} & \colhead{\B} & \colhead{\R} & \colhead{\ip} & \colhead{\z} & \colhead{\textit{NB}\tablenotemark{c}}
}
\startdata
SDSS J080715.2+132805	& 4.885 & 9.24 		& 47.07 & \nodata & 6000 & 4500 & 4200 & 9000 \\
&&&& \nodata & 1.1\arcsec & 1.0\arcsec & 1.3\arcsec & 0.9\arcsec \\
SDSS J111358.3+025333	& 4.870 & 9.12 		& 46.89 & \nodata & 3300 & 3900 & 3000 & 11700 \\
&&&& \nodata & 0.9\arcsec & 0.9\arcsec & 1.0\arcsec & 0.8\arcsec \\
4C 04.11				& 4.514 & \gtrsim9 	& \nodata & 4500 & 3600 & 5400 & \nodata & 24000 \\
(03{\mbox{$~\!\!^{\mathrm h}$}}11{\mbox{$~\!\!^{\mathrm m}$}}48{\mbox{$.\!\!^{\mathrm s}$}}0, +05\arcdeg08\arcmin01{\mbox{$.\!\!\arcsec$}}5)&&&& 0.9\arcsec & 0.8\arcsec & 0.9\arcsec & \nodata & 0.9\arcsec \\
\enddata
\tablenotetext{a}{From \citet{Trakhtenbrot2011} (J08 and J11) and \citet{Parijskij2014} (4C04).}
\tablenotetext{b}{Exposures during poor conditions were discarded.}
\tablenotetext{c}{\NBa for J08 and J11 fields and \NBb for 4C04 field.}
\end{deluxetable}
We conducted imaging observations of two fields around QSOs at $z\sim4.9$, 
SDSS J080715.2+132805 (08{\mbox{$~\!\!^{\mathrm h}$}}07{\mbox{$~\!\!^{\mathrm m}$}}15{\mbox{$.\!\!^{\mathrm s}$}}1, +13\arcdeg28\arcmin05{\mbox{$.\!\!\arcsec$}}2, 
$z=4.885$, \citealt{Trakhtenbrot2011}, hereafter J08 field) and 
SDSS J111358.3+025333 (11{\mbox{$~\!\!^{\mathrm h}$}}13{\mbox{$~\!\!^{\mathrm m}$}}58{\mbox{$.\!\!^{\mathrm s}$}}3 +02\arcdeg53\arcmin33{\mbox{$.\!\!\arcsec$}}7, 
$z=4.870$, \citealt{Trakhtenbrot2011}, hereafter J11 field) 
using the Suprime-Cam \citep{Miyazaki2002} on the 8.2m Subaru Telescope 
in 2014 Dec 21--26 and 2015 Dec 16--17 (UT; Program ID: S14B-006 and S15B-010, 
PI: M. Imanishi). As we illustrate in Figure \ref{fig:trans}, we 
used broad-band filters (\R, \ip, and \z) and a narrowband 
filter \NBa ($\lambda\_c = 7126$ \AA, $\Delta\lambda=73$ \AA) to sample LAEs 
in the redshift range of $4.83\lesssim z \lesssim4.89$ and LBGs at $z\sim5$. 
The redshifts of J08 and J11 are measured to be within this range 
in \cite{Trakhtenbrot2011} using the Mg {\sc ii} emission line, which is known 
to be good redshift indicator of type 1 AGNs \citep{Hewett2010,Shen2016b}. 
Note that, because their redshifts fall in the redder part of the 
sensitivity of \NBa and LAEs are more easily selected in bluer {\bf side} 
of the \NBa bandpass, there is a possibility that we detect LAEs at slightly 
different redshift. If true systemic redshifts of the AGNs are larger 
than the observed values, our measurements are affected further.
The median and intrinsic scatter of \ion{Mg}{2} redshift with 
respect to the narrow [\ion{O}{2}] line is measured 
to be $-$62 $\mathrm{km~s^{-1}}$ and 220 $\mathrm{km~s^{-1}}$,
respectively
\footnote{Also note that, in \citet{Trakhtenbrot2016b}, they detected [\ion{C}{2}] emission line in six AGNs in the \citet{Trakhtenbrot2011} sample with ALMA and found velocity shifts of $\sim$500 km s$^{-1}$ between the \ion{Mg}{2} line and the [\ion{C}{2}] line.} \citep{Shen2016b}.
Assuming Gaussian distribution, the probability of J08 having 
redshift higher than 4.89 is thus $\lesssim6$\%. 
$M\_{BH}$ of these two QSOs are also derived 
from the Mg {\sc ii} line and found to be very massive at $z\sim5$. 
Specifically, they have SMBHs with mass of $10^{9.24}$ M$_\odot$ for J08 and $10^{9.12}$ M$_\odot$ for J11. 
 
Additionally, we obtained S-Cam broad-band (\B, \R, and \ip) and narrow-band  (\NBb, $\lambda\_c = 6712$ \AA, $\Delta\lambda=120$ \AA) 
images of an RG at $z=4.514$ from the data archive (Program ID: S09B-070N, PI: Y. Matsuda). 
We can detect LAEs at $z=4.47$--4.57 with \NBb. 
This RG, namely 4C 04.11 (also known as RC J0311+0507; 
03{\mbox{$~\!\!^{\mathrm h}$}}11{\mbox{$~\!\!^{\mathrm m}$}}48{\mbox{$.\!\!^{\mathrm s}$}}0, 
+05\arcdeg08\arcmin01{\mbox{$.\!\!\arcsec$}}5 at
$z=4.514$, hereafter 4C04 field) has an estimated 
luminosity at rest-frame 500 MHz of 
$3\times10^{29}\mathrm{~W~Hz^{-1}}$, 
or luminosity at rest-frame 2.7 GHz of 
$L\_{2.7GHz}=6\times10^{34}\mathrm{~erg~s^{-1}~Hz^{-1}~sr^{-1}}$. 
For reference, the sample in \citealt{Venemans2007} has at most 
radio luminosity of 
$L\_{2.7GHz}=2.0\times10^{34}\mathrm{~erg~s^{-1}~Hz^{-1}~sr^{-1}}$. 
Its very high radio luminosity and a highly asymmetric Fanaroff-Riley 
type II (FR II) structure jet suggest that it is powered by 
a SMBH of mass $\sim10^9\mathrm{~M_\odot}$ \citep{Kopylov2006}.
The redshift of the RG is confirmed by various emission lines, such as 
the Ly$\alpha$, [\ion{O}{2}], and [\ion{Ne}{3}] lines \citep{Nesvadba2016a} and well constrained within $z=4.504$--4.514 
(see the inset of Figure \ref{fig:trans}).

At $z\sim5$, we can construct sufficiently large sample of galaxies with 
reasonable integration time. 
For J08 and J11 fields, the individual integration times in \R, \ip, \z, and \NBa 
were 300 sec, 300 sec, 300 sec, and 900 sec per pointing, respectively. 
For 4C04 fields, the individual integration times in \B, \R, \ip, and \NBb 
were 450 sec, 450 sec, 360 sec, and 1200 sec per pointing, respectively.
Each exposure was dithered by $\gtrsim60\arcsec$. 
A $N$ point circular dithering pattern ($N=6, 9$) was used. 
Details of our targets and observations are listed in Table \ref{tab:obs}.

The raw data were reduced in a standard manner with the dedicated software 
package, SDFRED2 
\citep{Ouchi2004}, which includes bias subtraction, flat-fielding, distortion 
correction, sky subtraction, image alignments, and stacking. 
The \NBa-band images were processed with L. A. Cosmic algorithm \citep{VanDokkum2001} 
to remove cosmic rays after the flat-fielding process. 
The world coordinate system of images were calibrated by comparing the 
USNO-B1.0 catalog. Mean offset from the catalog is $\sim0.2\arcsec$.
After we matched seeing of the images of each field to the worst one 
(see Section \ref{subsec:lae} and \ref{subsec:lbg}), we perform object detection 
and photometry using the double-image mode of SExtractor version 2.1.6 
\citep{Bertin1996}. 
We used the narrow-bands and the \z-band as detection bands for LAEs and LBGs, 
respectively. 
Spikes and halos around bright stars and low-S/N regions near 
the edge of FoV are masked during object detection. 
Objects in the masked regions or with SExtractor 
flags of $>2$ are eliminated from our catalogs. 
SExtractor flag 1 means an object has close neighbor or bad pixels 
which affect photometry. 
SExtractor flag 2 means an object was originally blended with another one. 
We first include these sources and later visually check and eliminate 
obvious spurious and heavily blended ones to maximize the detection rate.
 
We use magnitudes and colors measured in 1.7 times PSF FWHM diameter apertures 
unless otherwise stated. 
Photometric calibration for the J08 and J11 fields was obtained from the 
spectrophotometric standard stars, GD 108 \citep{Oke1990} and GD 153 \citep{Moehler2014}, 
observed during the same night of the observations. 
Photometric calibration for the 4C04 field was obtained from SDSS stars 
in the field.
We corrected the measured magnitudes 
for Galactic extinction of $E(B-V)=0.03$ mag (J08), 0.04 mag (J11), and 0.19 mag (4C04) 
\citep{Schlegel1998}. 
We note that the stellar colors measured in the J08 field are offset by $\sim$0.1--0.2 
mag from the \citet{Gunn1983} stellar templates on \R-\NBa vs. 
\NBa-\ip and \R-\ip vs. \ip-\z color-color diagrams. 
Since the \R- and \ip-band magnitudes of stellar sources in the field are 
confirmed to be well matched with those of SDSS sources, we manually corrected 
the zero-points of \NBa- and \z-band. 
The reason for this offset is unclear, 
but may partly be due to the unstable weather condition of our observations.  
Finally, the locus of stellar sources on a color-color diagram in all the fields 
became consistent with those of \cite{Gunn1983} within $\simeq$ 0.05 mag. 
Limiting magnitudes of our observations are given in 
Section \ref{subsec:lae} and Section \ref{subsec:lbg}.


\section{Analyses} \label{sec:sample}
\subsection{Selection of Lyman-alpha emitters} \label{subsec:lae}
As image quality of our \z-band image is not as good as we expected 
because of weather conditions and \z-band magnitude is not required for 
LAE selection below, we matched seeing of images of J08 and J11 field to that of 
\R-band (the worst one except for \z-band; see Table \ref{tab:obs}) 
when we construct LAE sample. 
The $3\sigma$ limiting magnitudes after all corrections in (\R, \ip, \NBa) are 
(27.5, 27.1, 26.6) and (27.5, 27.3, 26.7) for J08 and J11 field, respectively. 
We chose LAEs at $z\sim4.9$ from our catalog using the criteria below \citep{Ouchi2003}: 

\begin{eqnarray}
&\textit{Ri}_1 - \NBa > \max(0.8, 3\sigma(\textit{Ri}_1-\NBa))& \\
&\R - \ip > 0.5& \\
&\ip - \NBa > 0& \\
&\NBa < 5\sigma\_{NB711}, \ip < 2\sigma\_{i'},&
\end{eqnarray}
where $\textit{Ri}_1\equiv 0.5\times\R+0.5\times\ip$ and 
$\sigma$(\textit{Ri$_1-$}\NBa) denotes the expected deviation of the quantity 
\textit{Ri$_1-$}\NBa for a flat continuum source. 
For objects fainter than $2\sigma$ in \R-band, we replace the \R-band magnitude by
its $2\sigma$ limiting magnitude as a lower limit.
\citet{Ouchi2003} showed that these criteria effectively remove contaminants
such as low-redshift emitters and objects with a trough feature blueward of 
the \NBa-band filter.
\citet{Shimasaku2003} conducted follow-up spectroscopy of five candidates 
selected by this criteria and found a contamination rate of about $\sim20\%$, 
which is sufficiently low for our purpose. 

\begin{figure}
   \plotone{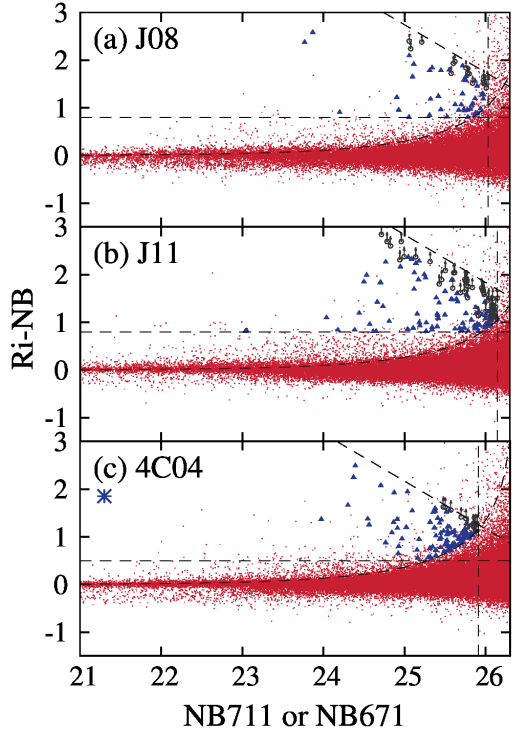}
 \caption{Distribution of sources in color-magnitude diagrams in (a) J08, (b) J11, 
and (c) 4C04 field. Red points show all sources in our catalog and blue triangles 
and gray circles with arrows show LAEs with and without \R-band detection ($2\sigma$). 
The asterisk in panel (c) shows the location of the RG 4C 04.11, which is also selected 
as LAE. The QSO J08 and J11 are also selected as LAE but they have \NBa$<21$. 
$x$-axis shows magnitude in \NBa-band in panel (a) and (b), \NBb in panel (c)) and 
$y$-axis shows $Ri_1-\NBa$ in (a) and (b), and $Ri_2-\NBb$ in (c). 
The dashed lines denote the criteria used to select LAEs (see Section \ref{subsec:lae}).
\label{fig:colormag}}
\end{figure}

In 4C04 field, $3\sigma$ limiting magnitudes in (\B, \R, \ip, \NBb) are 
(27.0, 26.8, 26.5, 26.5). We chose LAEs at $z\sim4.5$ using the criteria below: 
\begin{eqnarray}
&\textit{Ri}_2 - \NBb > \max(0.5, 3\sigma(\textit{Ri}_2-\NBb))& \\
&\ip - \NBb > 0& \\
&\NBb < 5\sigma\_{NB671}, \ip < 2\sigma\_{i'}, \textit{B} > 3\sigma\_B,&
\end{eqnarray}
where $\textit{Ri}_2\equiv 0.8\times\R+0.2\times\ip$ and 
$\sigma$(\textit{Ri$_2-$}\NBb) denotes the expected deviation of the quantity 
\textit{Ri$_2-$}\NBb for a flat continuum source. 
Similar criteria were used in Large-Area Lyman Alpha (LALA) survey 
\citep[e.g.][]{Rhoads2000} to detect LAEs at $z\sim4.5$ and 
the success rate based on their spectroscopic follow-up campaign is about 
$\sim70\%$ \citep{Dawson2007,Zheng2013}. 
In Figure \ref{fig:colormag} we show 
color-magnitude diagrams in each field. 

Finally, we visually checked the images and eliminated some spurious sources such as ones clearly blending with other sources and ones on stellar halos or saturation spikes. 
Finally, we find 60, 136, and 105 LAEs in J08, J11, and 4C04 field, respectively. 
Then we divide LAEs into two subgroups: one is ``proximity'' sample, which is 
located within 3 pMpc (J08 and J11 fields) or 5 pMpc (4C04 field) from the 
central AGNs, the other is ``blank fields'' sample, which is the rest of the 
sample. The size of the ``proximity'' of 3 or 5 pMpc is 
set by the FWHM of employed narrow-band filters ($\Delta \lambda\_{NB711} =72$ {\AA}, 
$\Delta \lambda_{NB671} =120$ {\AA}), and these values are sufficiently small 
to detect overdensities or galaxies affected by AGN feedback at $z\sim5$
\citep{Chiang2013,Calverley2011}. 
The sample size of proximity LAEs in J08, J11, and 4C04 field is respectively 14, 34, and 32.

\subsection{Selection of Lyman-break galaxies at $z\sim5$} \label{subsec:lbg}
\begin{figure}
   \plotone{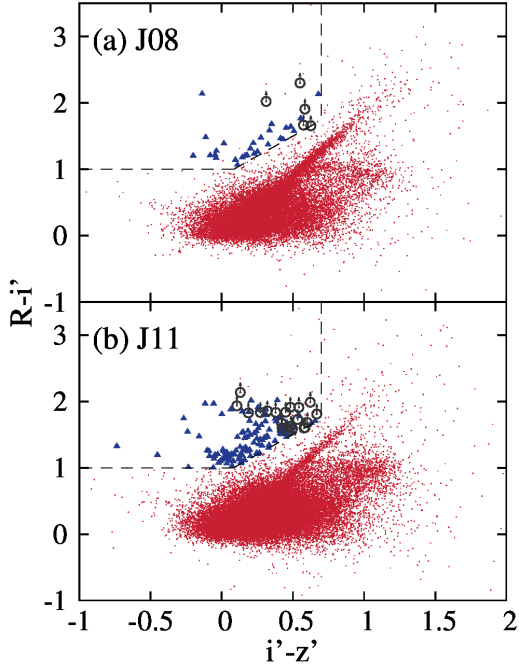}
\caption{Distribution of sources in color-color diagrams in (a) J08 and (b) J11 field. 
Red points indicate all sources with 5$\sigma$ detection in \z-band in our catalog,  
blue triangles and gray circles with arrows indicate LBGs with and without 
\R-band detection ($2\sigma$). 
The dashed lines denote the criteria used to select LBGs (see Section \ref{subsec:lbg}).
\label{fig:colorcolor}}
\end{figure}

When we matched seeing of images to that of \z-band, 
limiting $3\sigma$ aperture magnitude in (\R, \ip, \z) is (26.8, 27.3, 26.0) 
and (27.1, 27.3, 26.3) for J08 and J11 fields, respectively. 
We chose LBGs at $z\sim5$ from our catalog using the criteria below \citep{Yoshida2006}: 
\begin{eqnarray}
&\R - \ip > 1.0& \\
&\ip - \z < 0.7& \\
&\R - \ip > 1.2(\ip - \z) + 0.9& \\
&\z < 5\sigma\_{z'}.&
\end{eqnarray}
In Figure \ref{fig:colorcolor} we show color-color diagrams of J08 and J11 fields. 

In \citet{Yoshida2006}, the contamination rate of this {\it Ri'z'}-LBG sample is 
estimated via Monte Calro simulation based on photometric redshift catalog of galaxies in the Hubble Deep Field. 
The derived value of $\lesssim0.05$ is negligibly low. 
After visual inspection, the number of LBGs detected in J08 and J11 filed is respectively 33 and 137. 
LBGs are also divided into 
``proximity'' sample and ``blank fields'' sample as done for LAEs. 
The number of LBGs in proximity sample is 10 and 35 in J08 and J11 field, respectively.

\subsection{Physical Properties of LAEs and LBGs} \label{subsec:physpro}
We derived Ly$\alpha$ and UV luminosity and rest-frame equivalent width 
(EW$_0$) of LAEs assuming that the UV continuum slope is flat and
the \ip-band and narrow-band fluxes of LAEs are expressed as 
\begin{eqnarray}
f\_{i'}&=&f\_{cont}+F\_{Ly\alpha}/\Delta\_{i'} \nonumber \\
f\_{NB711}&=&f\_{cont}+F\_{Ly\alpha}/\Delta\_{NB711} \nonumber
\end{eqnarray}
in the J08 and J11 fields and 
\begin{eqnarray}
f\_{i'}&=&f\_{cont} \nonumber \\
f\_{NB671}&=&f\_{cont}+F\_{Ly\alpha}/\Delta\_{NB671} \nonumber
\end{eqnarray}
in 4C04 field, 
where $F\_{Ly\alpha}$ and $f\_{cont}$ respectively denote the Ly$\alpha$ flux in units of 
$\mathrm{erg~s^{-1}~cm^{-2}}$ and the continuum flux in units of $\mathrm{erg~s^{-1}~cm^{-2}~}$\AA$^{-1}$; 
$f\_{i'}$, $f\_{NB711}$, and $f\_{NB671}$ respectively denote \ip-, \NBa-, and \NBb-band flux 
in units of $\mathrm{erg~s^{-1}~cm^{-2}~}$\AA$^{-1}$; 
and $\Delta\_{i'}$, $\Delta\_{NB711}$, and $\Delta\_{NB671}$ respectively denote 
the FWHMs of \ip-, \NBa-, and \NBb-band in units of \AA. 
Then Ly$\alpha$ luminosity and EW$_0$ of LAEs are expressed by following formula:
\begin{eqnarray}
L\_{Ly\alpha} &=& 4\pi d\_L^2 F\_{Ly\alpha} \nonumber \\
&=&4\pi d\_L^2 \frac{\Delta\_{NB711}(f\_{NB711}-f\_{i'})}{1-\Delta\_{NB711}/\Delta\_{i'}} \label{eq:lyaforJ} \\
\text{or} ~~ &=& 4\pi d\_L^2 \Delta\_{NB671}(f\_{NB671}-f\_{i'}) \label{eq:lyafor4}
\end{eqnarray}
and
\begin{eqnarray}
\mathrm{EW_0}&=& F\_{Ly\alpha}/f\_{cont}(1+z) \nonumber \\
			&=& \frac{\Delta\_{NB711}(f\_{NB711}-f\_{i'})}{f\_{i'}-(\Delta\_{NB711}/\Delta\_{i'})f\_{NB711}}\frac{1}{1+z} \label{eq:EWforJ} \\
\text{or} ~~ &=& \frac{\Delta\_{NB671}(f\_{NB671}-f\_{i'})}{f\_{i'}}\frac{1}{1+z}. \label{eq:EWfor4}
\end{eqnarray}
Equation \ref{eq:lyaforJ} and \ref{eq:EWforJ} are used in J08 and J11 fields and 
Equation \ref{eq:lyafor4} and \ref{eq:EWfor4} are used in 4C04 field. UV luminosity of LBGs 
is directly derived from \z-band magnitude.

To derive number density and luminosity function of LAE and LBG, 
we have to calculate the effective volume surveyed and the completeness. 
These can be obtained by using Monte Carlo 
simulations in the same manner as done in many surveys such as \citet{Ouchi2003} 
and \citet{Yoshida2006}. 
For LAEs, we derived the effective volume by simply multiplying the surface area probed by 
the depth determined by the narrow-band filters we used. 
This gives total surveyed volume of $\sim1.5\times10^5$ cMpc$^3$ for J08 \& J11 fields 
and $2.9\times10^5$ cMpc$^3$ for 4C04 field
\footnote{These values include the masked regions. The masked regions are at most 10\% and 
thus negligibly small.}. 
When we calculate the LAE completeness, we used colors of LAEs 
with $>5\sigma$ detection in the \ip-band as artificial LAEs. 
We randomly distributed artificial point sources in the real images. 
Then we ran SExtractor and applied the same selection criteria to them as the real sample 
constructions. The completeness is defined as a ratio of the number 
of reproduced objects that again passed the criteria to the number 
of all of the input objects in unmasked regions. 
Completeness ($C\_{NB}(m\_{NB})$) correction was done by weighting 
a LAE which has NB magnitude $m\_{NB}$ by $1/C\_{NB}(m\_{NB})$. 
This simple method gives sufficiently reliable estimates \citep{Shimasaku2006}. 
Note that our main purpose here is to compare the relative shape of luminosity functions 
in and out of the proximity and not to compare the absolute values of them. 
On the other hand, the effective volume and completeness of LBGs cannot be obtained 
immediately. 
We derive the effective surveyed volume $V\_{eff}(m)$ as follows:
\begin{eqnarray}
V\_{eff}(m)	&=& \int_0^\infty C(m,z) \frac{dV\_C(z)}{dz}dz \nonumber \\
		&=& \int_0^\infty C(m,z) \frac{c}{H_0 E(z)} d\_c^2 d\Omega dz
\end{eqnarray}
where $C(m,z)$ is the completeness of a LBG at redshift $z$ with apparent 
magnitude of $m$; 
$V\_C(z)$ is a comoving volume at redshift $z$; $d\_C$ is 
comoving distance; and 
$E(z)\equiv\sqrt{\Omega\_m (1+z)^3 + \Omega\_\Lambda}$.
The completeness $C(m,z)$ is 
calculated in a similar way to \citet{Yoshida2006}: we generate 
mock LBGs using the stellar population synthesis model 
of \citet{Bruzual2003} with varying dust extinction 
($E(B-V)=0.0$--$0.5$, $\Delta E(B-V)=0.1$, adopting Calzetti's extinction 
law \citep{Calzetti2000}). The distribution of $E(B-V)$ values is taken from that 
of $z\sim4$ LBGs measured in \citet{Ouchi2004}. 
Age and metallicity are kept to the constant value of 0.1 Gyr and 0.2$Z_\odot$, 
respectively. We used an exponentially decaying star formation history with 
the e-folding time of 5 Gyr. After redshifting the spectra 
to $z=4.4$--$5.3$, we corrected for the effect of intergalactic absorption 
by neutral hydrogen by adopting the model of \citet{Madau1995}. Then we 
derive the colors of the model LBGs and input them as point sources into 
the real images. Source detection and photometry were done in the same manner 
as in the real sample constructions. Finally, we derived the completeness 
as functions of \z-band magnitude and redshift.

\section{Results and Discussion} \label{sec:results}
\subsection{Environments of high-redshift AGNs} \label{subsec:env}
\begin{figure*}[ht!]
\plotone{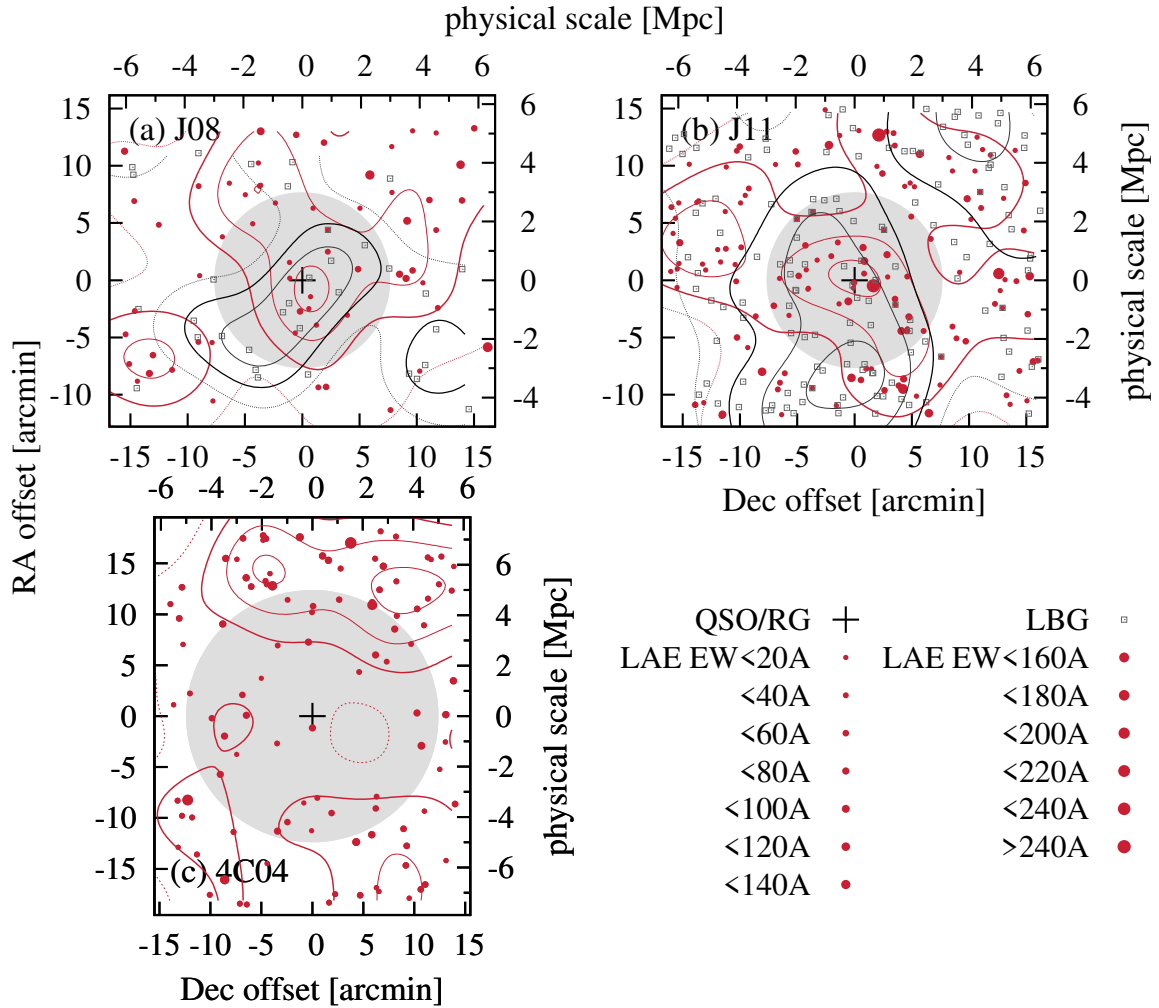}
\caption{Sky distribution of LAEs in (a) J08, (b) J11, and (c) 4C04 field. 
The plus signs at the center denote the location of the AGNs. 
Filled red circles denote LAEs. The size of the circles depends on Ly$\alpha$ EW$_0$. 
Filled gray squares denote LBGs. 
The gray shaded areas show the proximity of 3 pMpc (J08 and J11 fields) or 
5 pMpc (RG field) centered at each AGN. 
Red and black contours indicate surface density $\delta\_{gal}\equiv(n-\bar n)/\bar n$ of LAEs and LBGs 
(including QSO/RG) from 1.0 to -0.5 with a step of 0.5 in the case of J08 and 4C04 and 0.8 to -0.8 with a step of 0.4 in the case of J11, 
derived by the fixed aperture method; 
we distributed apertures of radius 8 cMpc and counted the number of galaxies within them 
($\equiv n$), and derived the average ($\equiv\bar n$) in each field. 
Thick solid lines indicate $\delta\_{gal}=0$ and dotted lines show negative $\delta\_{gal}$. 
Note, however, that number density at the edge of each field is underestimated and contours are plotted just to guide the eye.
\label{fig:maps}}
\end{figure*}

We show in Figure \ref{fig:maps} the sky 
distributions of the central AGNs, LAEs and LBGs (J08 \& J11 fields only) 
selected in Section \ref{subsec:lae} and \ref{subsec:lbg}. 
The two QSOs are located close to the local peaks of surface density, 
while the RG is isolated. 
The significance of the peaks near the QSOs is lower than $2\sigma$ in either case. 
This level of variance is found in other blank field surveys  
\citep{Ouchi2003,Ouchi2005,Shioya2009}. 
\citet{Chiang2013} derived median and 1$\sigma$ scatter of overdensities 
of galaxies with SFR $>$ 1 $M_\odot/$yr, 
$\delta\_{gal}\equiv(n-\bar n )/\bar n$ (see their Figure 13 for $z=3$ case) 
as a function of $\Delta z$ by using a semi-analytic model. 
In our cases $\delta\_{gal}$ is at most $\sim$1. 
Note that the area of the window of 15 $\times$ 15 cMpc used in \citet{Chiang2013} is close to our aperture size (circle of 8 cMpc radius). 
At least they are not likely to be 
associated with the most massive overdensity at $z\sim5$ or 
evolve into ``Coma'' type clusters with $M\_{halo}>10^{15}M_\odot$, 
though there remains the possibility of them being ``Fornax'' type cluster 
($M\_{halo}<3\times10^{14}M_\odot$). 
Though the density peak of LBGs in J08 field is also near J08 itself, 
the distribution of LBGs as a whole is clearly different from that of LAEs.

\begin{figure*}
\plotone{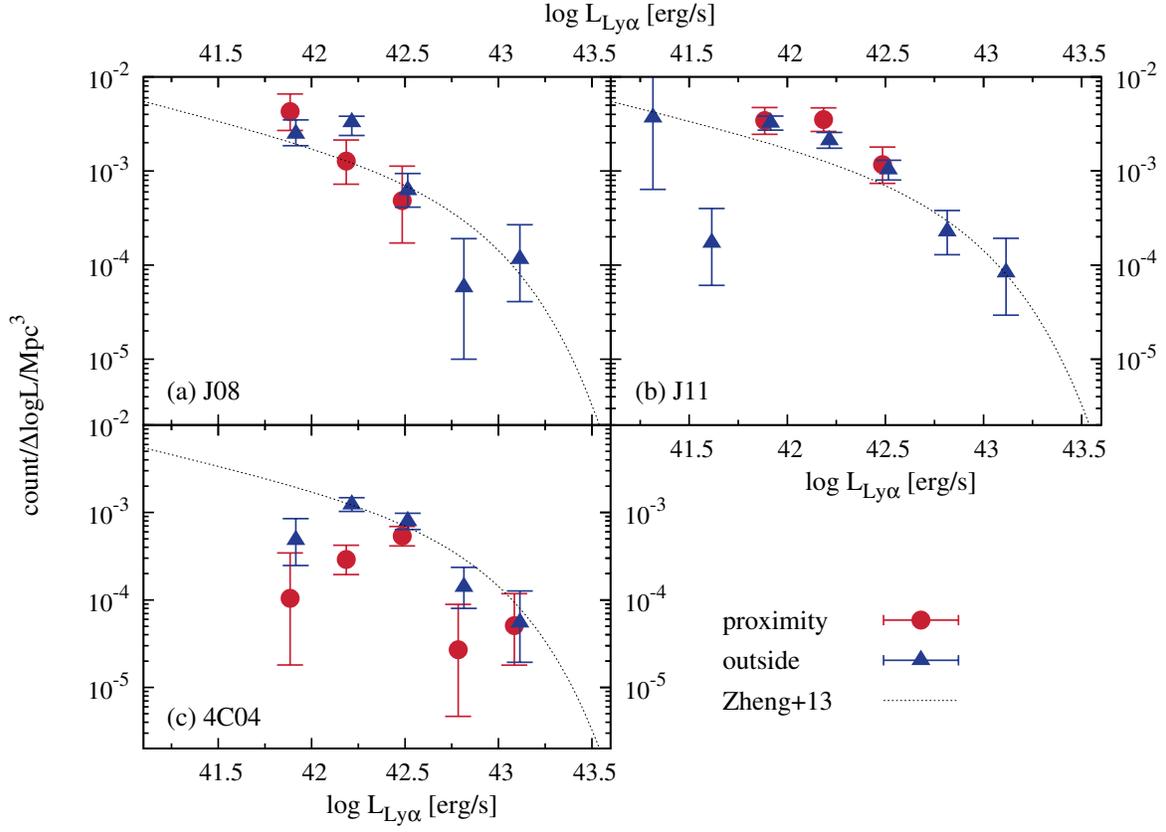}
\caption{Ly$\alpha$ luminosity functions of LAEs in (a) J08, (b) J11, and (c) 4C04 
field. Red circles and blue triangles respectively represent LFs 
in the proximity and in the outer region. The dotted line is a LF of LAE at $z\sim4.5$
derived in \citet{Zheng2013} using spectroscopically confirmed LAEs with $L_\mathrm{Ly\alpha}>10^{42.5}$.
\label{fig:lyalf}}
\end{figure*}

\begin{figure*}
\plotone{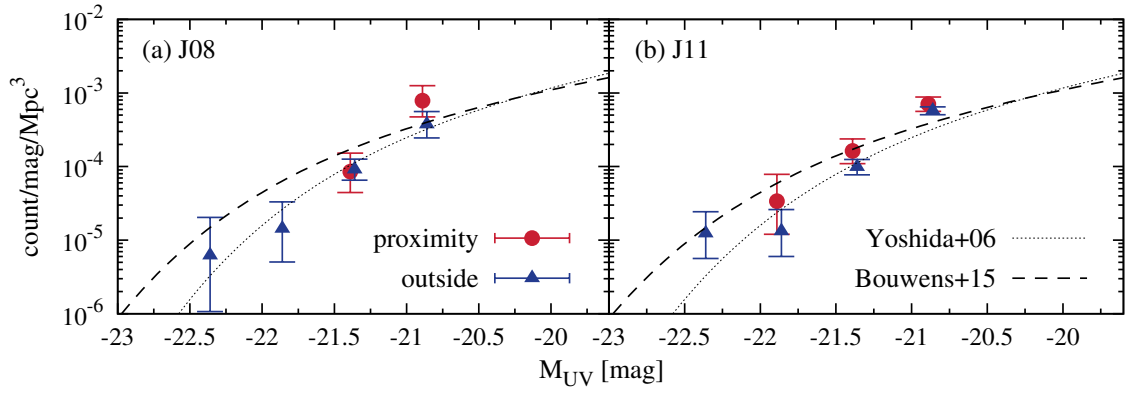}
\caption{UV luminosity functions of LBGs in (a) J08 and (b) J11 field. 
Red circles and blue triangles respectively represent LFs 
in the proximity and in the outer region. 
The dotted and dashed lines are respectively the LFs at $z\sim5$ derived in 
\citet{Yoshida2006} and \citet{Bouwens2015}. These papers use different selection criteria to select galaxies at $z\sim5$. Our selection criteria are the same as those used in \citet{Yoshida2006}.
\label{fig:lbglf}}
\end{figure*}

The trend of no massive overdensity becomes clearer in Figure \ref{fig:lyalf} and \ref{fig:lbglf}, in which 
we respectively show the Ly$\alpha$ luminosity functions (LFs) of LAEs and 
UV LFs of LBGs in each field. 
The error bars represent 84\% single-sided confidence levels based on the Poisson statistics 
\citep{Gehrels1986} alone. 
We find no significant excess of LFs in the proximities 
compared to that in blank fields. 
Rather, we find no LAEs in the two brightest bins in the proximities of 
the two QSOs at $z\sim4.9$. We also find no LBG in the two brightest bins and the brightest bin in J08 and J11 field, respectively. 
This is reasonable considering 
the small volume probed and the smaller number density of brighter 
LAEs and LBGs. Indeed, if we assume that the number of these galaxies follows 
the Poisson distribution and that the surface density 
of galaxies in the proximity are the same as that in 
the outer region, the probability of finding 
no LAEs with $L_\mathrm{Ly\alpha} > 10^{42.75}$ 
(and LBGs with $M\_{UV}<-21.6$) 
in the proximity of J08 and J11 is 36\% and 14\% 
(36\% and 10\%), respectively. 
These threshold values correspond to those of the second brightest bin in the LFs in Figure \ref{fig:lyalf} and \ref{fig:lbglf}. 
On the other hand, if we assume the surface density of LAEs (LBGs) 
in the QSO proximity is twice as high as blank fields, probabilities of non-detection
decrease to 13\% and 1.9\% (13\% and 1.1\%), respectively.
As we discuss further in section \ref{subsec:feedback}, 
the impact of AGN feedback seems to be negligible in our data;
even at the fainter side, there is no significant difference. 
This trend holds true for the case of LBGs (Figure \ref{fig:lbglf}), 
which suggests that these QSOs do not reside in overdensities 
of much larger scales considering large $\Delta z$ of 
the LBG selection function. 

\begin{figure}
\plotone{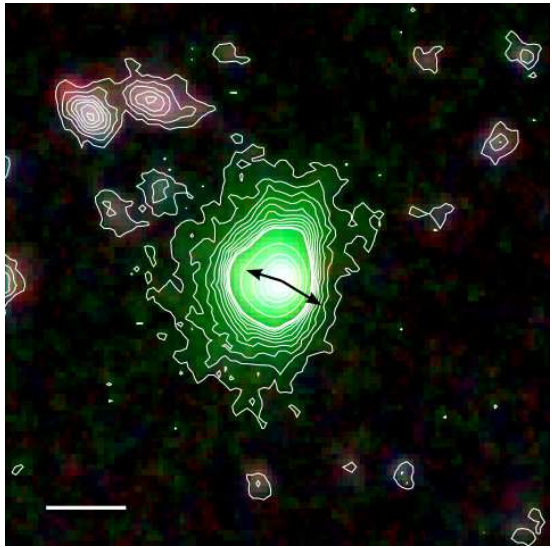}
\caption{Pseudo-color image of 4C04 with north to right and east to up. 
R, G, and B correspond to \ip-, \NBb-, and \R-band, respectively. 
Contour level is different above and below the thick line ($6.1\times10^{-17}\mathrm{~erg~s^{-1}~cm^{-2}~arcsec^{-2}}$) for clarity: 
the step size outside of the thick line corresponds to
$6.1\times10^{-18}\mathrm{~erg~s^{-1}~cm^{-2}~arcsec^{-2}}$ and the step size 
inside the thick line is 20 times larger than outside. 
The scale bar at the bottom left of the figure indicates 20 pkpc ($\sim$ 3\arcsec). 
The size of the image is 20\arcsec$\times$20\arcsec. 
The black arrows indicate the direction of a radio jet and the location 
of bright lobes in \citet{Parijskij2014}.
\label{fig:4C04}}
\end{figure}

The RG 4C04 is found to be near a void. The number density of LAEs in the proximity of the RG is somewhat 
lower than in the outer region. This is at odds with the results in the 
literature \citep{Zheng2006,Ajiki2006,Venemans2007}, in which 
the authors report overdensity of LAEs around radio-loud QSOs and RGs 
at various redshifts with a high success rate ($\gtrsim70$\%). 
We note that 4C04 has extremely luminous ($L_\mathrm{Ly\alpha}>10^{44} \mathrm{~erg~s^{-1}}$) 
and extended ($\gtrsim$ 60 kpc) Ly$\alpha$ halo around it (Figure \ref{fig:4C04}). 
High redshift RGs (HzRGs) often have such extended Ly$\alpha$ halos. 
They usually align with their radio jets \citep{Venemans2007,Nesvadba2016a}. 
However, the 4C04 halo is rather extended almost perpendicular to the jet 
direction (arrows in Figure \ref{fig:4C04}). 
The Ly$\alpha$ nebula extending far beyond the radio emitting region, and 
the non-detection of overdensity and the \ion{C}{4} and \ion{He}{2} 
lines \citep{Kopylov2006} which are frequently detected in HzRGs with line ratios suggestive of an enriched outflow origin, all 
suggest that this RG is a different class of objects from other HzRGs. 
Additionally, though the upper limit of the line ratios of the halo are consistent with bright 
(type-1) QSOs \citep{Borisova2016}, the moderately broad Ly$\alpha$ line width 
($\sim$1500 km/s, \citealt{Kopylov2006}) is not consistent with those QSOs. 
Further observations are needed to conclude 
the origin of this Ly$\alpha$ nebula.

Considering a large scatter in the number of galaxies even with Suprime-Cam FoV 
\citep[can be as large as $\sim0.3$ depending on the bias factor of LAEs, e.g.][]{Shimasaku2004,Gawiser2007,Trenti2008}, it is hard to draw a firm conclusion from our 
observations alone. Nonetheless, observing more and more AGN fields with small $\Delta z$ 
and comparing the results with theories greatly help us understand the true nature 
of AGN environments. 
Besides our results, there is growing observational evidence that high-redshift 
($z\gtrsim5$) AGNs do not reside in overdensities when one probes them 
with narrow-band filters ($\Delta z\lesssim0.1$) at least 
on a scale smaller than 3 pMpc
(\citealt{Banados2013,Mazzucchelli2016} at $z=5.7$). 
These results suggest that an overdensity on this scale (and possibly galaxy merger) 
is not always a necessary condition for AGN activity. 
Some semi-analytic models predict that halos of high-redshift 
AGNs are not the most massive ones at that epoch \citep{Overzier2009,Fanidakis2013,Orsi2016}. 
Recently \citet{DiMatteo2016} suggested that 
the most massive BHs at the earliest epoch ($z>8$) can be preferentially 
formed in regions where tidal fields are weak because gas can directly 
fall onto BHs along filaments, rather than mere overdense regions. 
On the other hand, \citet{Costa2014b} argued that SMBHs of $10^9~\mathrm{M_\odot}$ 
at $z\sim6$ only formed in the most massive halos as a result of their cosmological 
hydrodynamical simulations. 
The environments identified in theoretical works depend largely on their assumptions 
about BH seeding, BH growth, and AGN feedback. For example, 
models usually assume that AGNs are triggered only by galaxy-galaxy mergers. 
Furthermore, we have to rely on simple subgrid physics because nuclear regions
are impossible to resolve in galaxy scale simulations. 
Clearly models need to be tested and updated with recent observational indications. 
Although AGN luminosity is still important in terms of AGN feedback, 
it is also crucial to investigate the AGN environment as a function of BH mass. This is because AGN luminosity is rather an 
instantaneous, time-changing (``differential'') physical quantity determined by SMBH mass and accretion rate at 
the observed time. On the other hand, BH mass is a more fundamental ``cumulative'' one which reflects its growth history and thus more likely to be related to its large scale environments. 

Finally, we mention another possibility: we failed to trace the environments with LAEs and LBGs 
because, for example, most of galaxies around AGNs may be dusty and UV/optically elusive. 
Many lines of evidence suggest that LAEs are typically young ($\lesssim100$Myr), 
low-mass ($\lesssim10^8\mathrm{~M_\odot}$) and non-dusty ($E(B-V)\lesssim 0.2$) 
galaxies \citep{Gawiser2006,Ono2010a,Ono2010,Finkelstein2007,Finkelstein2009} 
with some massive and dusty outliers. As there are many systems with 
a large amount of dust even at this high redshift \citep{Riechers2013}, 
we could largely miss such a population. 
Recently, \citet{Trakhtenbrot2016b} observed six $z\sim4.8$ QSOs from the same parent sample 
we used with ALMA and detected companion SMGs in three of the observed 
AGNs. High probability of finding companion SMGs in a small FoV of ALMA and 
faintness of the companions in other wavelengths (not detected even in the Spitzer data) 
indicate that there may be many optically-elusive galaxies around AGNs in our fields. 
Since the techniques utilizing UV/optical feature 
are biased against dust obscured galaxies, wide-field (far-)infrared 
observations are needed to complement such techniques. 

\subsection{Feedback from AGNs} \label{subsec:feedback}
It has long been theoretically argued that ultraviolet background (UVB) radiation 
can suppress the formation of low-mass galaxies by heating their gas 
\citep{Efstathiou1992,Thoul1996,Kitayama2000,Kitayama2001,Benson2002a,Dijkstra2004,
Susa2004,Mesinger2008,Okamoto2008,sobacchi2013a,sobacchi2013b,Liu2016a}. 
In principle, we can confirm this suppression by identifying 
flattening of the faint-end slope of the LF, 
though observationally this turned out to be extremely difficult 
\citep{Weisz2014,Alavi2014,Alavi2016a,Castellano2016}; 
no evidence of an LF turnover has been found down to $M\_{UV}\sim-12$. 
On the other hand, there are a handful of candidate ultra-faint dwarf galaxies 
at the local Universe 
\citep[][$M\_V>-8$ or $M_*\lesssim10^4 ~\mathrm{M_\odot}$]{Brown2014} whose 
star-formation seem to have been suppressed by a synchronized external process 
such as the reionization. 
This kind of negative feedback is often provoked to explain why we do not always 
find high-redshift QSOs in overdensities because local UV radiation fields around 
luminous QSOs are much stronger than the UVB.
Previous studies originally considered situations where a pregalactic cloud is 
collapsing under UVB in the reionization era 
and studied its evolution as a function of various parameters such as the intensity 
of UVB, the time at which UVB is switched on and the pregalactic cloud starts to collapse 
\citep[see also][]{Kashikawa2007}. 
As a result, they derived the threshold dynamical 
mass below which a pregalactic cloud cannot collapse and form stars. The results showed 
a range of the threshold mass of $\sim10^9$--$10^{10}~\mathrm{M_\odot}$, 
depending on the details of calculations (e.g., 1D/3D, including radiative transfer 
effect or not.). 
They also found that once the clouds begin to collapse and their density get higher, 
even very strong UV radiation cannot affect the evolution of the clouds and 
finally allows the clouds to form stars, whereas 
clouds which are irradiated well before they start to collapse can be affected significantly 
\citep{Kashikawa2007,sobacchi2013b,Roos2015}. 
In many cases the strength of UVB is kept constant, but in some cases gradually evolving UVB is 
assumed. The impact of UVB feedback is maximized when clouds are irradiated by constant UVB 
before they begin to collapse.

Although the assumptions in the above theoretical studies are not exactly 
matched to the situation studied in this paper (i.e., UVB vs. the AGN proximity), 
we can refer to these results to infer in what circumstances the impact of 
radiative AGN feedback becomes significant. 
First, a semi-analytic model predict halo masses of faint 
($10^{41}<L\_{Ly\alpha}<10^{42}\mathrm{~erg~s^{-1}}$) LAEs,
moderate-luminosity ($10^{42}<L\_{Ly\alpha}<10^{43}\mathrm{~erg~s^{-1}}$) LAEs, and 
LGBs with UV magnitude brighter than $M\_{UV}=-20.8$ at $z\sim5$ of 
$\sim10^{10.4}~\mathrm{M_\odot}$, $10^{11.1}~\mathrm{M_\odot}$, and 
$10^{11.7}~\mathrm{M_\odot}$, respectively \citep{Garel2015}. 
Halo mass of faint LAE is in agreement with the threshold mass. 
Second, the strength of UV radiation field at 3 pMpc from our QSOs 
is stronger than that of UVB at $z\sim5$. It is parametrized as 
$J(\nu) = J_{21} \times (\nu/\nu_0)^\alpha \times10^{-21} 
\mathrm{~erg~s^{-1}~Hz^{-1}~cm^{-2}~sr^{-1}}$, where $\nu_0$ denotes the Lyman limit 
frequency and $\alpha$ denotes continuum slope. We derive $J_{21}$ at 3 pMpc from the QSOs 
assuming UV continuum slope $\alpha=-0.99$ \citep{Fan2001} and using the measured luminosity 
at 1450{\AA}, $L_\mathrm{1450}$ from \citet{Trakhtenbrot2011}, to find 
$J_{21}=1.2$ and $0.7$ for J08 and J11, respectively. 
Inferred UVB radiation $J_{21}$ at $z\sim5$ is of order 0.1 \citep{Calverley2011} 
and is similar to the assumed value in previous calculations which predict 
strong feedback. Thus, taken at face value, 
fainter LAEs in our sample can be significantly affected by those QSOs. 

However, there are some caveats in the above arguments. 
First, in order to suppress the formation of LAEs significantly, QSOs should be 
switched on before LAEs around them starts to form. 
This can be the case if the lifetime of the QSO phase is similar to an estimated 
maximum value of $\sim10^8$--$10^9$ yr \citep{Marconi2004}, 
since estimated age of LAEs is $\sim$ a few $\times1$--$100$ Myr \citep{Finkelstein2009}.
Second, the short time-scale variability of the UV source is 
not taken into account in the theories. 
QSOs show strong variability on timescales as short as days. At the same time, 
many simulations showed that AGNs change their luminosity dramatically in the course of 
major mergers and also there are some indications of AGN flickering on timescales 
$\sim10^5$ yr \citep{Schawinski2015}. 
If the variability is taken into account, the impact of 
AGN feedback is further weakened because high-redshift AGNs probed so far are 
considered to be near its peak luminosity and its time-averaged luminosity may be lower 
\citep{Hopkins2006,Hopkins2009}.
Third, although primordial gas is assumed in the calculations, 
pregalactic clouds can contain metals and dust grains 
which do not originate from in situ star formation but from neighboring galaxies 
via various mechanisms such as galactic winds and radiation pressure. 
This makes cooling and heating processes more complex: metals 
\citep{Wiersma2009} and dust emission contribute 
cooling once the temperature and density get high enough ($>10^6$ K) and 
make star formation easier, whereas 
photoelectric dust heating could be quite efficient and negative feedback could be stronger 
at the temperature, UVB strength, and density condition of the earlier stage of collapse 
considered here \citep{Nath1998,Montier2004}. 

Though the effect of metals and dusts is unknown, 
if a QSO activates well before surrounding galaxies start to collapse, 
and retains its high luminosity for $10^8$--$10^9$ years, 
the suppression of faint galaxies due to a QSO can be observable. 
These conditions may not be always fulfilled (e.g., there are many works 
claiming an episodic and short QSO phase; see \citealt{Martini2003}). 
Even if it exists, only LAEs fainter than 
$L\_{Ly\alpha}=10^{42} \mathrm{~erg~s^{-1}}$ and LBGs fainter than 
$M\_{UV}=-18.3$ seem to be critically affected. 
Since limiting magnitudes of observations conducted in the past are much brighter, 
it is difficult to explain the deficit of overdensity around QSOs reported so far 
by radiative feedback, unless halo masses of LAEs are overestimated. 
More realistic calculations and much deeper observations are clearly needed to 
qualitatively examine AGN radiative feedback.

\subsection{Fluorescent emission} \label{subsec:fluorescent}
\begin{figure}
   \plotone{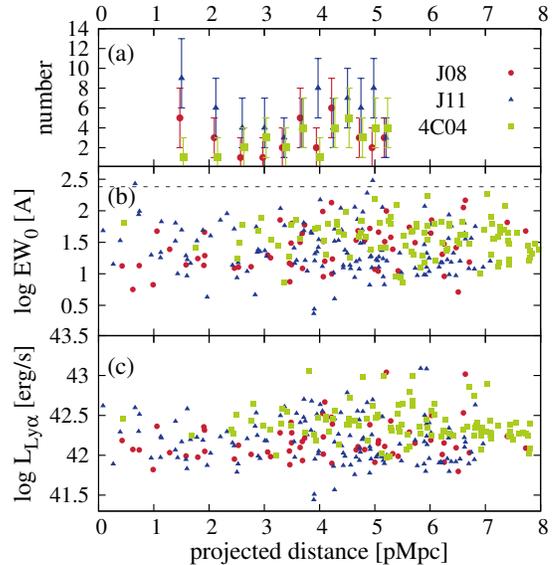}
 \caption{Properties of LAEs as a function of the projected distance from the central AGNs. 
(a): The number of LAEs with $L\_{Ly\alpha}>10^{42} \mathrm{~erg~s^{-1}}$ in annuli 
centered at J08 (Red circles), J11 (Blue triangles), 
and 4C04 (Green squares). Widths of annuli was chosen so that each annulus has the same area. 
(b): Rest-frame equivalent width. A horizontal dashed line represent EW of 240 \AA. 
(c): Ly$\alpha$ luminosity of LAEs.
\label{fig:radial}}
\end{figure}

High-EW LAEs are interesting because such high 
Ly$\alpha$ EW ($>240$ \AA) is unlikely to be due solely to normal star formation
\citep{Charlot1993,Schaerer2003a}. 
It is usually attributed to clumpy, dusty IGM 
\citep{Hansen2006,Finkelstein2008,Kobayashi2010} or 
fluorescent Ly$\alpha$ emission \citep{Adelberger2006,Cantalupo2005,Cantalupo2007} 
especially in the case of QSO environments. 
\citet{Cantalupo2012} reported that many Ly$\alpha$ fluorescent systems illuminated by a 
hyperluminous QSO at $z=2.4$ are clustered around the QSO and argued that, as opposed 
to the case of feedback (Section \ref{subsec:feedback}), the fainter-side of 
Ly$\alpha$ LF of LAEs around the QSO becomes steeper due to these sources. 
In our sample, only two LAEs in J11 field have such high EW. 
One of the LAEs is located at {\bf $\sim0.7$ pMpc} (projected) from 
J11. The other is located at $\sim5$ pMpc (projected) from 
J11. Though the former can be significantly affected by radiation from 
the QSO, the latter are unlikely to be significantly affected, since at that distance QSO radiation is comparable to UVB. 
\citet{Cantalupo2012} predicted that if fluorescence is dominant 
in the field, $L\_{Ly\alpha}$ of LAEs should decrease with increasing distance 
from the QSO. 
Figure \ref{fig:radial} shows the number, Ly$\alpha$ equivalent widths (EWs) and 
Ly$\alpha$ luminosities of LAEs as a function of the projected distance from 
the central AGNs. In Panel (a), we see the possible signature of local peak 
seen in Figure \ref{fig:maps}, i.e., 
the rising trend of the number of LAEs at $<2$ pMpc in J08 and J11 field. 
However, in Panel (b) and (c), we find no dependence on the distance in any field. 
This suggests that the properties of most of the LAEs in these fields are not affected by 
the central AGNs. 

\begin{figure}[ht!]
\plotone{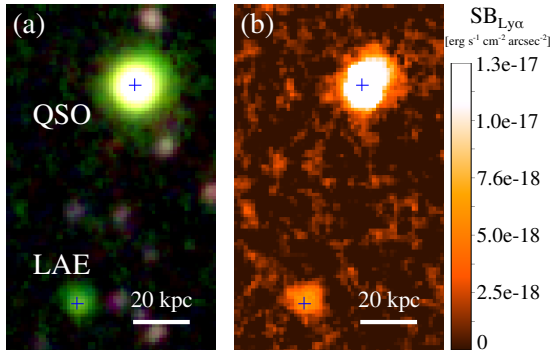}
\caption{Pseudo-color image of J11 and the close companion marked respectively as ``QSO'' and ``LAE'' (Panel (a)) and 
Ly$\alpha$ image (Panel (b)). 
The scale bar at bottom right of the figure indicates 20 pkpc 
($\sim$ 3\arcsec). 
The size of this panel is 20\arcsec$\times$12\arcsec. 
The plus signs show the location of the \ip-band peak of each source.
In Panel (a), R, G, and B correspond to \ip-, \NBa-, and \R-band, 
respectively. 
In Panel (b), extended Ly$\alpha$ halos around both sources can be seen.
\label{fig:J11companion}}
\end{figure}

There is one interesting source in the J11 field: a close companion LAE located at 
80 pkpc away from J11. 
We show the pseudo-color image {\bf and the continuum-subtracted Ly$\alpha$ image} of J11 and 
the companion in Figure \ref{fig:J11companion}. 
{\bf The companion clearly shows a Ly$\alpha$ halo which extends toward the QSO direction} and may suggest interactions between these two galaxies. 
Though $\Delta z$ of \NBa filter is large compared to 80 pkpc, 
if the companion is at the same redshift as J11, $J_{21}$ will be $\sim1000$. 
Following \citet{Cantalupo2005}, we derived the ``effective boost factor'' 
$b\_{eff}$ to be $\sim3000$. Thus if there exist optically thick clouds, they can be 
detected with our imaging observations. 
Ly$\alpha$ luminosity of the QSO near side is $\sim$1.8 times that of the far side. 
Thus, although EW is not high (EW$_0\sim48$ \AA), fluorescent emission may contribute 
to its $L\_{Ly\alpha}$, if not all. 
We also note that the QSO J11 itself showed asymmetric, extended Ly$\alpha$ emission 
toward the opposite side of the companion. This may imply a giant filamentary 
structure passing through these two galaxies. 
Confirming this issue needs further observations and will be discussed elsewhere. 

\section{Summary and conclusions} \label{sec:summary} 
High-redshift AGNs are thought to be signposts of highly biased regions of the Universe. 
Previous studies on their environments though presented confusing results, 
suggesting that AGNs reside in both rich and poor environments. 
This is partly due to the use of different techniques and possibly radiative feedback 
from AGNs. The conventional Lyman-break technique only gives us an unclear picture 
because it selects galaxies from wide redshift range ($\Delta z\sim1$, corresponds to 
$\sim100$ physical Mpc at the redshift of interest of this study, $z\sim5$), 
making it difficult to discuss local (a few-several physical Mpc, corresponds to 
$\Delta z \sim 0.1$) environments around AGNs. 
In order to test whether AGN environments are rich and whether AGN feedback is 
indeed strong enough to suppress formation of neighboring galaxies, 
we conducted deep and wide-field imaging observations with the Suprime-Cam 
on the Subaru telescope and searched for LAEs around two QSOs at $z\sim4.9$ and 
an RG at $z\sim4.5$ by using narrow-band filters. 
In QSO fields, we also obtained additional broad-band images to select 
LBGs at $z\sim5$ for comparison. 
We constructed a photometric sample of 301 LAEs and 170 LBGs in total. 
A wide field of view ($34\arcmin\times27\arcmin$) of the Suprime-Cam 
enabled us to probe these galaxies in the immediate vicinities of the AGNs 
and in the blank fields simultaneously and compare various properties of them 
in a consistent manner. 
We find that the QSOs are located near low peaks of galaxy surface density, 
though the data suggest they are not uncommon (with $<2\sigma$ significance), 
and one of the QSOs has a close companion LAE with projected distance $\sim80$ physical kpc.
However, the luminosity functions of LAE/LBG around the QSOs and RG are consistent with 
or lower than those in blank fields as opposed to the expectation that 
they should reside in the most massive overdensities. 
Moreover, we find no evidence of feedback even in the faintest luminosity bin 
(down to $L_\mathrm{Ly\alpha} =10^{41.8} \mathrm{~erg~s^{-1}}$). 

Through our discussion in Section \ref{subsec:feedback}, 
we conclude that radiative feedback is unlikely to affect our sample 
and galaxies around high-redshift AGNs observed to date. 
Therefore our results suggest that high-redshift AGNs do not necessarily trace 
overdense regions of the Universe and that is not due to radiative feedback. 
Note that most of the currently known QSO fields with significant overdensity of 
neighboring galaxies are detected with LBGs, and that 
spectroscopic follow-up of them is very challenging with the existing instruments. 
Thus there still remains a possibility that 
the photometric sample around those overdense regions is significantly 
affected by the projection effect. Further observations with narrow-band filters 
around high-redshift AGNs are the best way to know the true
nature of their environments. 
In parallel, observations around them with submillimeter/millimeter facilities like ALMA \citep{Trakhtenbrot2016b} is crucial for detecting possible dusty galaxies. 
Large-area surveys will find more and more AGNs at high-redshift with their 
redshifted Ly$\alpha$ emission line falling into the existing narrow-band filters.
For example, many narrow-band filters are installed on 
the Hyper Suprime-Cam \citep{2012SPIE.8446E..0ZM}, which has much wider FoV (1.5 degrees 
in diameter) than the Suprime-Cam and thus is the most powerful instrument for 
this study. Particularly, Subaru High-z Exploration of Low-Luminosity Quasars 
\citep[SHELLQs;][]{Matsuoka2016} will find many high-redshift low-luminosity QSOs, 
including QSOs with very massive SMBHs but with low accretion rate, and will help us 
reveal more general trends of high-redshift AGN environments. 

\acknowledgments
We greatly appreciate the anonymous referee for his/her thoughtful comments.
We thank Tomonori Totani and Nobunari Kashikawa for helpful comments and discussions. 
We would like to appreciate all the staffs who supported our observations with the Subaru Telescope, including the staffs of the National Astronomical Observatory of Japan, Mauna Kea Observatory, and local Hawaiian people who have been making effort to preserve and share the beautiful dark sky of Mauna Kea with us. 
MI is supported by JSPS KAKENHI Grant Number 15K05030. 
YM is supported by JSPS KAKENHI Grant Number 17H04831.
KS is supported by JSPS KAKENHI Grant Number 16K05286.

\vspace{5mm}
\facilities{Subaru(Suprime-Cam)}

\software{SDFRED2, L.A.Cosmic, SExtractor, IRAF}

\end{document}